\documentclass[10pt,global,twocolumn]{svjour}
\usepackage{times,mathptmx}
\usepackage{graphicx, subfigure}
\usepackage{url} 
\usepackage{amsmath}
\usepackage{dcolumn}
\usepackage{amssymb}
\usepackage[super,negative]{nth}
\newcommand{\co}[2]{#1 \tiny{$\pm#2$}} 

\begin{document}
\title{The heterogeneity of inter-contact time distributions:\\its importance for routing in delay tolerant networks}

\author{Vania Conan\inst{1}, J\'er\'emie Leguay\inst{1}\inst{2}, Timur Friedman\inst{2}}

\institute{
  Thales Communications\and
  Universit\'e Pierre et Marie Curie, Laboratoire LIP6--CNRS
}

\maketitle

\begin{abstract}
  Prior work on routing in delay tolerant networks (DTNs) has commonly
  made the assumption that each pair of nodes shares the same
  inter-contact time distribution as every other pair. The main
  argument in this paper is that researchers should also be looking at
  heterogeneous inter-contact time distributions. We demonstrate the
  presence of such heterogeneity in the often-used Dartmouth Wi-Fi
  data set. 
  We also show that
  DTN routing can benefit from knowing these distributions. We first
  introduce a new stochastic model focusing on the inter-contact
  time distributions between all pairs of nodes, which we validate on
  real connectivity patterns. We then analytically derive the mean delivery
  time for a bundle of information traversing the network for simple
  single copy routing schemes. The purpose is to examine the
  theoretic impact of heterogeneous inter-contact time distributions.
  Finally, we show that we can exploit this user diversity to improve
  routing performance. 
\end{abstract}

\section{Introduction}

In the kind of delay tolerant networks (DTNs)~\cite{dtn_fall_sigcomm}
that we consider in this paper, nodes are
mobile and have wireless networking capabilities. They are able to
communicate with each other only when they are within transmission
range. The network suffers from frequent connectivity
disruptions, making the topology only intermittently and partially
connected. This means that there is a very low probability that an
end-to-end path exists between a given pair of nodes at a given time.
Such DTNs can be consider in ad hoc networking when connectivity
is very low (e.g. in tactical military communications), in 
transportation systems as in the DieselNet project~\cite{UMassDieselNet} or in 
Pocket Switched Networks (PSN)~\cite{chaintreau} which are formed by devices that 
people carry everyday (cell phones, PDAs, music players).
In all these contexts, end-to-end paths can exist temporarily, or may sometimes never exist,
with only partial paths emerging. This paper addresses the extreme
case, where only temporal paths exist. We call such networks
\textit{temporal DTNs}, or \textit{t-DTNs}.
When a node in a t-DTN receives a ``bundle'' of information from a 
neighboring node, it keeps it until it meets another node which 
provides an opportunity to relay the bundle. The bundle is transfered 
from one node to another instantly and this transfer is atomic.

Prior work on routing in t-DTNs has commonly made the assumption that
each pair of nodes shares the same inter-contact time distribution as
every other pair.  The main argument in this paper is that researchers
should also be looking at cases in which inter-contact time
distributions are heterogeneous. Chaintreau et al.~\cite{chaintreau} posit that there might be
heterogeneity, but we show it and characterize it. We also show how
exponential distributions can be composed to yield the heavy-tailed
distributions that Chaintreau et al.\ observed.  As we shall see, the
heterogeneity that we highlight allows us to usefully extend the work
of Spyropoulos et al.~\cite{spyro04single,spyro04multi}, which
analyzes numerous routing schemes for t-DTNs, but that uses mobility
models that yield homogeneous distributions.

We show, on the well known Dartmouth Wi-Fi data set~\cite{dartmount},
that despite the existence of a heavy-tailed distribution when
inter-contact times are considered in the aggregate, a large portion
of the node pairs present inter-contact time distributions that can be
well fitted by an exponential distribution. We found these
distributions to be heterogeneous, with a wide variation in exponents.

We also provide the first formal analysis of the impact of
heterogeneous exponential inter-contact time distributions on simple
single-copy routing schemes. We show that routing strategies can
benefit, in terms of delay, from this heterogeneity, and in particular
from knowing these distributions.  A node can choose among possible
relay nodes based upon their expectations for meeting other relays or
the destination.

\section{Inter-contact time model}\label{sec:model}

This section presents the model we use to analytically derive
the delay expectations for the routing protocols we study
later in this paper.

\subsection{Exponential t-DTNs}

We consider a network composed of \textit{n} nodes. Let's first look
at the inter-contact time between two individual nodes $(i,j)$:
$t_{ij}^1 < t_{ij}^2 < t_{ij}^3 < ...$ are the successive instants at
which a contact between $i$ and $j$ occurs.
\begin{equation}
\label{eq_tau}
\Delta t_{ij}^k=t_{ij}^{k+1} - t_{ij}^k
\end{equation}
is the inter-contact time between the $k^{th}$ and $(k+1)^{th}$ contact instants.

We assume that the $\Delta t_{ij}^k$ are samples from independent and
identically distributed random variables that follow an exponential
law with parameter $\lambda_{ij}$, which we note $\tau_{ij}=\mathrm{exponential}(\lambda_{ij})$.
The mean inter-contact time between $i$ and $j$ is thus given by $E[\tau_{ij}]=
1/\lambda_{ij}$.

In the overall network, all $n$ nodes are supposed to behave independently, 
so that the $n(n-1)/2$ pairwise inter-contact times $\tau_{ij}$ are independent 
exponential processes with different parameters. The $\tau_{ij}$ family of 
processes is symmetric and $\forall i, \tau_{ii} = 0$.
The exponential t-DTN is thus entirely and uniquely characterized by the $n(n-1)/2$ strictly 
positive real parameters $\lambda_{ij}$.

\subsection{Assumptions}

The model focuses on the temporal dynamics of node connectivity in a DTN. In this way it
provides a common framework to analyse different DTNs. In
particular it applies very well to social networks for which the
position of nodes at a given time is not of primary importance. We
believe this abstraction helps focus on the inherent characteristics
of intermittent connectivity in DTNs.

Characterizing inter-contact time behavior
helps abstract away from the spatial information that is
essential in the analysis of mobile ad hoc networks. There is no
reference to geographic, localisation or any other such spatial
information. There is also no reference to air interface parameters,
quality of or contention on the links, etc. Node mobility is not
explicitly modelled: only its aggregated impact on the inter-contact
time is taken into account.

The model makes a stationarity hypothesis with respect to node
inter-contact time distributions. In other words, nodes behaviors are
assumed to change on a slower scale than bundle exchanges.
We also suppose that nodes have infinite capacity in bandwidth and storage. 

Another key hypothesis is that contacts (and thus bundle transfers) are assumed to be instantaneous.
In the model, pairwise contacts do not overlap: in contrast to the mobile ad hoc network cases, no partial routes (involving more than two nodes) exist at any given time.

In this respect, the results with the proposed model are
upper bounds, but, as we will see, still provide valuable information
and insight on how to route bundles in DTNs. We leave refinements of
the model for future work.

\section{Fitting the model}\label{sec:fitting}

In the t-DTN model just elaborated, we assume that the inter-contact
time distribution for each pair of nodes is exponential. The main
reason is that it will allow us to go beyond asymptotic results and
provide explicit formulas for the bundle delivery time, and other
parameters, of different routing protocols. In this section we look
at real data to evaluate how reasonable this hypothesis might be.

\subsection{Experimental data set}

To validate the hypothesis, we use real data from
the Wi-Fi access network of Dartmouth College~\cite{dartmount}. 
These data track users' sessions in the wireless network by 
showing the time at which a node associates or dissociates from an
access point. We use the subset of data pre-processed by Song et al.\ for their 
prior work~\cite{song:predict} on mobility prediction.

As we describe in prior work~\cite{leguay06}, we must select from the
data, and make some assumptions, in order to constitute a useful DTN
data set. We take the subset of users who are present in the
network every day between January \nth{26} 2004 and March \nth{11}
2004, a class period during which we expect nodes' activity to be fairly
stationary. This data set contains 834 users, or nodes.  Then, we
assume that two nodes are in contact if they are present at the same
time at the same access point (AP). Finally, we filter these data to
remove the well known \emph{ping-pong} effect. Indeed, wireless nodes, even
non-mobile, can oscillate at a high frequency between two APs. 
To counter this, we filter all the inter-contact times below 1,800 seconds. 
Note that defining better filtering methods, albeit challenging, would 
be of interest for the community. As this is not the purpose of this work, 
We choose here the threshold that Yoon et al.~\cite{yoon06} used 
for the same purpose. We use this new data set for the remainder of this paper. 

The Wi-Fi scenario may be not a perfect fit for interactions between nodes in t-DTNs.
Indeed, in opposition to always-on devices carried by humans, Wi-Fi nodes are typically
turned off, transported, and then turned on
again, thus missing potential contacts en route. However, the size,
quality, and public availability of the data set make it nonetheless
one of the best resources for this kind of study.
Jones et al.~\cite{jones_wdtn} and Chaintreau et al.~\cite{chaintreau}
recently used these traces in a similar way.

\subsection{Exponential inter-contacts}

\begin{figure}[!h]
\centering
\includegraphics[width=5cm]{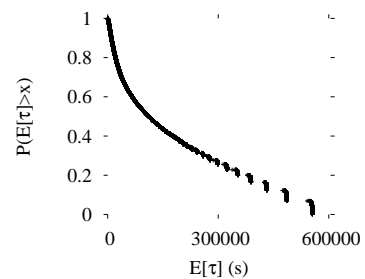}
\caption{\label{lambda_dist}Distribution of $E[\tau]$.}
\end{figure}

Fig.~\ref{lambda_dist} shows the distribution of $E[\tau_{ij}]$, 
the expected inter-contact time for the pair of nodes $(i,j)$.
This plot has been computed over
all the $28,490$ source-destination pairs that experienced an average inter-contact 
time lower than one week within the two months
period that we considered in Dartmouth data.
We can see that the distributions are heterogeneous with expectations varying
over three orders of magnitude. The average $E[\tau]$ is $11.6$ hours with a variance of 7.1 hours.

Then, we test for whether the inter-contact process between any two nodes
can be modelled by an exponential process with a parameter
$\lambda=1/E[\tau]$. We use
the Cramer-Smirnov-Von-Mises~\cite{Eadie71} hypothesis test.  For each
pair $(i,j)$, we compare the cumulative distribution $I_N^{ij}$ for
the $N$ inter-contacts observed and the hypothesis function whose
cumulative distribution is
$F_{ij}(x)=1-\mathrm{exp}(-\lambda_{ij}x)$. We also compare $I_N^{ij}$
with that of a power law distribution. Note that we only perform the
computation for pairs that show a sufficient level of connectivity by
having a mean inter-contact time lower than one week and that have
more than $20$ contacts. We identify 8,402 pairs to be exponentially
distributed and $28$ with a power law which makes respectively,
$62.3$\% and $0.2$\% of the $13,482$ pairs that we retain for the test. 

From these observations, it seems clearly more reasonable, in this data set, to model pairwise
inter-contact time distributions as exponential rather than power law 
since a large number of pairs have shown inter-contact times
exponentially distributed. Despite that few are power law distributed, we conjecture that
the rest of pairs might follow distributions that are a mix
of exponential and power law distributions.
As we have examined only one data set, albeit an often-used one, we
cannot draw many conclusions about what will be revealed elsewhere.
It is reasonable to expect that other mobility traces in campus
environments will show similar characteristics. However, it is
surprising that a memoryless process seems to be at work in such a
high proportion of node pairs in an environment in which one would
expect some temporal correlations.  We hope this will be a spur to
study these distributions in other data sets.
Traces from the Haggle project~\cite{chaintreau} or 
the ones of the Reality Mining project~\cite{realitymining} might be considered.
We let this study for future work.

\subsection{Power laws}

\begin{figure}[!h]
\centering
\includegraphics[width=5cm]{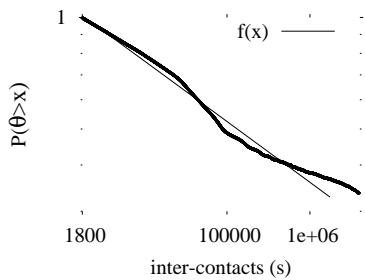}
\caption{\label{interc_dist}Distribution of inter-contacts in logarithmic scale.}
\end{figure}

Chaintreau et al.~\cite{chaintreau} observed that aggregated
inter-contact times follow power laws in a number of DTN 
traces (also including one based on the Dartmouth data).
Fig.~\ref{interc_dist} shows that, for our data set, the cumulative
distribution of aggregated inter-contact times also follows a power
law of the form $f(x)=cx^{\delta}$, with exponent $\delta=-0.16$ and
scale parameter $c=3.45$.

Let's now consider what happens for pure exponential t-DTNs where all
pairwise inter-contact time distributions are exponentially
distributed. Under which conditions do the aggregated inter-contact
time distributions follow a power law, or is the pairwise exponential
assumption too strong to yield a power law in the aggregate?

Let $\Theta$ be the aggregated inter-contact time for all pairs of
nodes, and let $p(\lambda)$ be the probability distribution of the
$\lambda$ parameters:

\begin{equation}
\label{eq_laplace}
P(\Theta>t)= \int_{\lambda=0}^{\infty}{e^{-\lambda t} p(\lambda) d\lambda}
\end{equation}

What eqn.~\ref{eq_laplace} says is that, for exponential t-DTNs, the 
aggregated inter-contact time distribution is fully characterized by 
the distributions of the $\lambda$ parameters, and thus of the $E[\tau_{ij}]$ 
matrix. More precisely, the tail cumulative distribution of the 
aggregated inter-contact times is given by the Laplace transform of 
the distribution $p$ of the $\lambda$ parameters.

A Pareto law of the form $(\frac{b}{t+b})^{\alpha}$, $t \geq 0$, with shape 
parameter $\alpha>0$ and scale parameter $b>0$, is observed if and 
only if the $\lambda$ follow a gamma 
distribution $p(\lambda)=\frac{\lambda^{\alpha-1}b^\alpha e^{-b\lambda}}{\Gamma(\alpha)}$, $\lambda \geq 0$. 

\begin{figure}[!h]
\centering
\subfigure[$\lambda_{ij}$]{\label{lambda_th}\includegraphics[width=3.8cm]{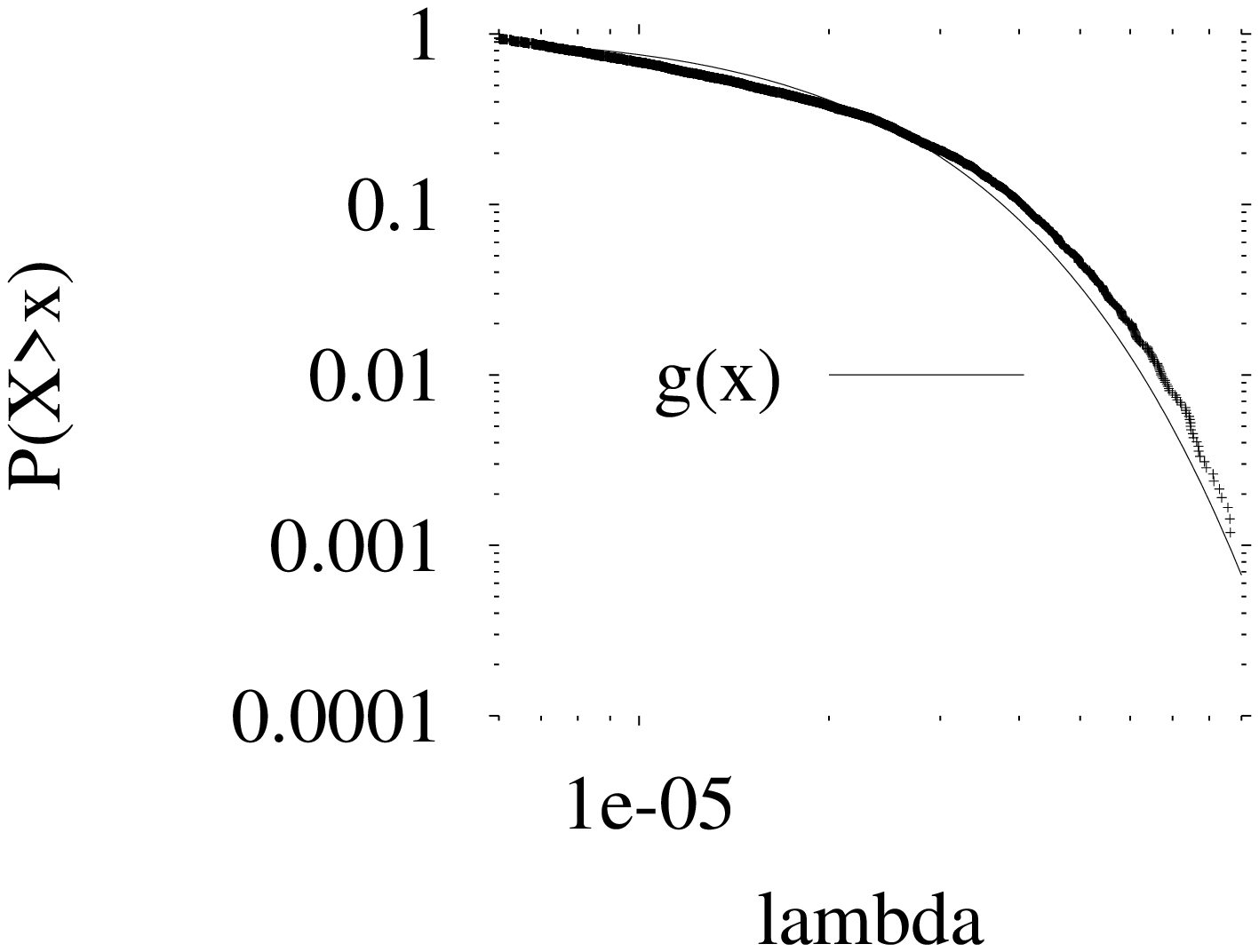}}
\subfigure[Inter contacts]{\label{interc_th}\includegraphics[width=3.8cm]{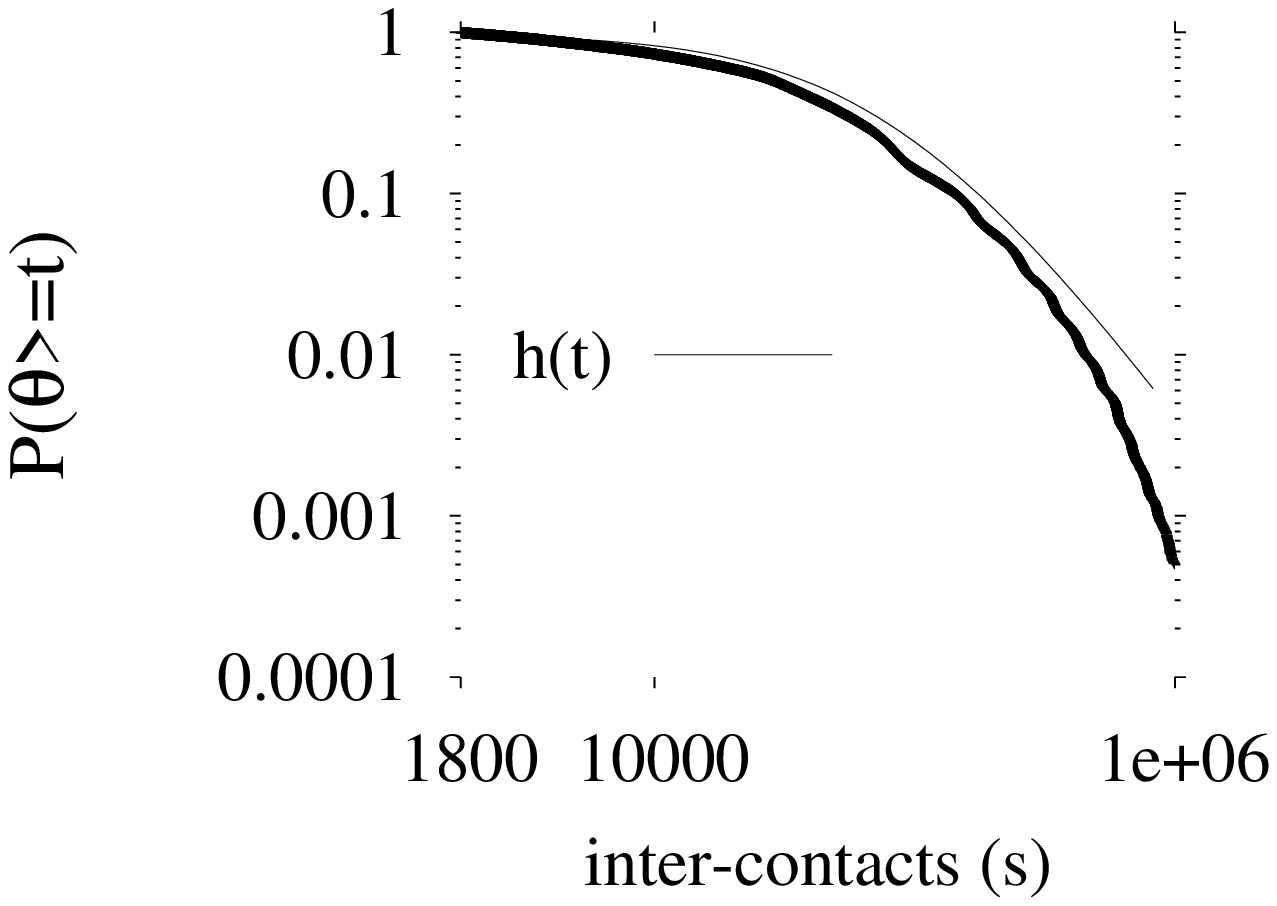}}
\caption{\label{th}Distributions with exponential pairs.}
\end{figure}

To verify this on the data set we proceed in the following way:
we estimate parameters $\alpha$ and $b$ from the cumulative distribution of the $\lambda$ 
parameters for pairs that were shown to follow an exponential behavior 
(the ones that pass the Cramer hypothesis test). We find $b=113,766.9$ 
and $\alpha=2.26$. Fig.~\ref{lambda_th} shows
the estimated cumulative gamma distribution $g(x)$ with the experimental lambda 
cumulative distribution for all pairs that have shown to be exponential.
Then, we plot in Fig.~\ref{interc_th} the corresponding power-law 
$h(t)$ with cumulative distribution of aggregated inter-contact times. 
As one can see, the two experimental curves fit the theoretical curves.

What this result shows is that when one considers an exponential t-DTN, we 
can regain the power law behavior for the aggregated inter-contacts when 
the distribution of the parameters is a gamma, which is the case in the data we 
used when considering the subset of pairs that have inter-contact times 
exponentially distributed. 

\section{Single copy routing strategies}\label{sec:formal}

Having defined a stochastic model that is realistic for the data set
under study, we now examine different simple single copy routing
strategies. We derive analytical formulas that we will use to study the
impact of heterogeneous $\lambda_{ij}$ parameters on routing.

In all routing strategies, we consider that nodes know all pairwise
mean inter-contact times for all nodes in the network, i.e., each node
knows the $\lambda_{ij}$ matrix. This knowledge could be diffused
through an epidemic type of routing, or learned by each node from past
contacts.

\subsection{Wait scheme}
\label{sec:Wait}

Under the Wait routing strategy, the
source node $s$ waits until it meets $d$, the destination, to
deliver the bundle in one hop.

If the bundle is injected at time $t$, its delivery time is equal to
$R_{sd}^{t}$, the remaining inter-contact time before the next contact
between nodes $s$ and $d$. The memoryless nature of exponentials
implies that $R_{sd}^{t}$ also follows an exponential distribution
with the same parameter.  The mean expected delivery is thus given by:

\begin{equation}
\label{eq_Dwait}
E[D^w_{sd}]= 1/\lambda_{sd}
\end{equation}

This straightforward result gives an upper bound on the delivery time that a 
routing strategy should meet, since the Wait strategy is the most rudimentary 
one hop single copy scheme.

\subsection{MED}

The Minimum Expected Delay (\textsc{Med}) routing strategy was first
introduced by Jain et al.~\cite{routingdtn}.  This strategy, similar
to source routing, defines which path the bundle will follow from $s$
to $d$, that is, the ordered list of intermediate relay nodes it will
have to go through.  The list is chosen to provide minimum expected
end-to-end delay.

If a path is given by the following ordered list of nodes $r_0=s < r_1
< r_2 < r_3 < ... < r_{n-1} < r_n = d$, and relaying occurs at time
instants $t_1 < t_2 < ...  < t_n$, the total delivery time along path
$(s,r_1,r_2, ..., r_{n-1},d)$ is given by the remaining inter-contact
time after each relaying instant, that is:
\begin{equation}
D^{med}_{s,r_1,r_2, ..., r_{n-1},d}= R_{sr_1}^{t_1}+R_{r_1r_2}^{t_2}+ ... + R_{r_{n-1}d}^{t_n}
\end{equation}

Using the fact that $E[R_{r_ir_j}]= 1/\lambda_{ij}$, the expected delivery time along the path is thus given by:
\begin{equation}
\label{eq_Dos}
E[D^{med}_{s,r_1,r_2, ..., r_{n-1},d}]= 1/\lambda_{sr_1}+1/\lambda_{r_1r_2}+...1/\lambda_{r_{n-1}d}
\end{equation}

Finding the optimal path thus amounts to finding a lowest-weight path
between nodes $s$ and $d$ in a graph in which the weight on each link
$(i,j)$ is defined as $1/\lambda_{ij}$.  Dijkstra's algorithm can be
used.

\subsection{Spray and Wait routing}
\label{sec_SW}

The Spray and Wait strategy was first introduced by Grossglauser and
Tse~\cite{GrossglauserTse02}, and is designed to take advantage of
opportunistic contacts.  It consists of two steps. First the source
node uses the first nodes encountered as relays to the destination.
This is the ``spraying'' step. A relay node then uses the ``wait''
strategy to relay the bundle, i.e. it waits until it meets the
destination to deliver the bundle.  Here, we study the case where only
one relay is used, which we designate 1-SW.

Let us first consider the spraying step. The bundle is injected at
source $s$ at time instant $t$. The first node $r$ it encounters may
be any of the $n-1$ other nodes $d, r_1, r_2, ..., r_{n-2}$ and the
time $X$ it takes to meet this first node is the infinum of the
inter-contact times with all other nodes:
\begin{equation}
X=inf(R_{sd}^{t},R_{sr_1}^{t}, ..., R_{sr_{n-2}}^{t})
\end{equation}
Since all $R_{sr_i}^{t}$ are independent exponentials with parameters
$\lambda_{sr_i}$, we have (see~\cite[p.328]{Bremaud99}):
\begin{itemize}
\item The random index $r$ of the first node encountered is
  independent of the first encounter time $X$
\item $X$ is exponentially distributed, with parameter: \\ $\Lambda_s=
  \lambda_{sd} + \sum_{i=1}^{n-2} (\lambda_{sr_i})$
\item Pr($First\ node\ encountered\ is\ r$) $=
  \frac{\lambda_{sr}}{\Lambda_s}$
\end{itemize}

This means that we can represent the spraying step as independently
identifying the encountered node (with probability
$\frac{\lambda_{sr}}{\Lambda_s}$) and adding an exponential waiting
time with parameter $\Lambda_s$.

Two cases may arise: either the first node encountered $r$ equals $d$,
and $s$ delivers the bundle, or $r \neq d$ and node $r$ waits to meet
node $d$ to deliver the bundle.

The delivery time $Z_d$, when node $d$ is encountered first is thus given by:
\begin{equation}
E[Z_d]= \frac{1}{\Lambda_s}
\end{equation}

The delivery time $Z_r$ along path $r$, i.e., conditioned on using
node $r$ as a relay, is thus the sum of the first encounter time $X$
and the remaining delivery time between nodes $r$ and $d$, and thus:
\begin{equation}
E[Z_r]= \frac{1}{\Lambda_s} + \frac{1}{\lambda_{rd}}
\end{equation}

The total delivery time $Z$ is computed by conditioning on all
possible first encountered nodes $d, r_1, r_2, ..., r_{n-2}$, events
whose probabilities are given by $\frac{\lambda_{sr}}{\Lambda_s}$:

\begin{equation}
E[Z]=\frac{\lambda_{sd}}{\Lambda_s} E[Z_d] + \sum_{i=1}^{n-2} ( \frac{\lambda_{sr_i}}{\Lambda_s} E[Z_{r_i}])
\end{equation}

After simplification, we can state that: in a network
composed of $n$ nodes, 1-SW delivers a bundle
from source $s$ to destination $d$ with mean delivery time given by:
\begin{equation}
\label{eq_Dsw}
E[D^{1-sw}_{sd}]= \frac{ ( 1 + \sum_{r \neq s, r \neq d} \frac{\lambda_{sr}}{\lambda_{rd}}) } {\sum_{r \neq s} \lambda_{sr}}
\end{equation}

\section{Comparing routing protocols}\label{sec:comp}

This section looks at routing performance of the protocols we considered in 
the presence of heterogeneity in inter-contact time distributions.

In this context, we present 1-SW$^*$, a variation of 1-SW.  Instead of
spraying its bundle to the first node that it encounters, the source
node $s$ sprays only to nodes in a subset $R$. We call this a
1-SW$^R$ scheme. 
We define 1-SW$^*$ to be a 1-SW$^R$ scheme which uses a subset
$R$ that minimizes $E[D^{1-sw^R}_{sd}]$.
Following the same line of reasoning as in
Sec.~\ref{sec_SW}, and defining $1/\lambda_{dd}=0$, one finds that the
expected delivery time is given by:
\begin{equation}
\label{eq_SHOSW}
E[D^{1-sw^R}_{sd}]= \frac{ ( 1 + \sum_{r \in R} \frac{\lambda_{sr}}{\lambda_{rd}}) } {\sum_{r \in R} \lambda_{sr}}
\end{equation}

We performed simulations using Dartmouth traces
(see Sec.~\ref{sec:fitting}) to study how the algorithms 
behave in the case of heterogeneous connectivity. We simulate the
following protocols: Wait and 1-SW which are naive schemes, and, 1-SW$^*$
and \textsc{Med} that are designed to take advantage of heterogeneity. 
We slightly modified 1-SW, to better compare it with
1-SW$^*$: a node $i$ is a potential relay only if $\lambda_{id}>0$,
i.e., if it has a chance of meeting the destination. In \textsc{Med},
we authorized intermediary relays to directly transfer bundles to the 
destination whenever met.

We choose at random $100$ different source destination pairs $(s,d)$ 
and replay the contacts between the $835$ nodes present in the data to see
how for each pair a bundle, generated at the beginning of the two months period, 
is delivered.

$\lambda$ values used for route selection in
1-SW$^*$ and \textsc{Med}, and to
determine theoretical delays of 1-SW, 1-SW$^*$ and \textsc{Med} have been computed
over the data filtered due to the ping-pong effect (see Sec.\ref{sec:fitting}).
However, the contacts replayed in simulations were that in the original
traces as it does not impact the results and because filtering was only
of interest for modelling.

\begin{table}[!h]
\setlength{\tabcolsep}{2pt}
\begin{center}
\small
\begin{tabular}{|c|c|c|c|c|c|c|c|c|}
\hline
  & \textbf{delivery} & \textbf{A delay} & \textbf{M delay} &\textbf{th. delay} & \textbf{hop count} \\
  & \textbf{ratio} (\%) & (days) & (days) & (days) & (hops)\\
\hline
\textbf{Wait} & \co{11.2}{0.9} & \co{19.8}{3.7} & \co{16.3}{10.3} & \co{41.3}{0.5} & \co{1.0}{0.0} \\
\textbf{1-SW} & \co{87.3}{3.0} & \co{23.0}{0.9} & \co{22.7}{2.7} & \co{15.3}{0.9} & \co{2.0}{0.1} \\
\textbf{1-SW$^*$} & \co{86.9}{2.0} & \co{18.8}{0.4} & \co{15.4}{0.9} & \co{13.0}{1.1} & \co{2.0}{0.1} \\
\textbf{MED} & \co{87.9}{2.2} & \co{20.9}{0.7} & \co{18.0}{1.1} & \co{1.3}{0.2} & \co{7.2}{0.2} \\
\hline
\end{tabular}
\end{center}
\caption{\label{res_simus}Simulation results with Dartmouth data.}
\end{table}

Table~\ref{res_simus} presents the simulation results averaged over $5$
runs with the $90$\% confidence levels that are obtained using the
Student $t$ distribution. It presents, for each of the protocols, the
average delivery ratio, the average delay (``A delay'') and the median
delay (``M delay'') computed over the delivered bundles, the average
theoretical delay over all the bundles generated (infinite delay is assumed to 
be the length of the simulated period, i.e. $45$ days), and the average 
hop count, also obtained on delivered bundles.  

The major result in Table~\ref{res_simus} is that schemes that make use of heterogeneity of inter-contact times (1-SW$^*$ and \textsc{Med}) perform better, either in delivery ratio or delay, than the ones that
do not exploit it (Wait and 1-SW). Wait only delivers 11.2\% of bundles
because most of the source, destination pairs selected at random
satisfy $\lambda_{sd}=0$ (e.g. they never met). 1-SW, 1-SW$^*$
and \textsc{Med} achieve almost the same delivery ratios with respectively 
87.3\%, 86.9\% and 87.9\%. About 13\% of the bundles were thus not delivered.
In terms of delay, among these last three protocols, 1-SW plots the highest
with a mean of 23.0 and a median of 22.7 in days, 
1-SW$^*$ the lowest with a mean of 18.8 and a median of 15.4. \textsc{Med}
appears to be in the mid-range with a mean of 20.9 and a median of 18.0 in days.

The difference between the modified 1-SW and 1-SW$^*$ gives a further insight on the type of heterogeneity
that should be considered. The modified 1-SW is a one hop strategy that uses only true relays to the destination: relay nodes in 1-SW must meet both the source and the destination. The scheme is not completely ignorant of heterogeneity, as it exploits binary connectivity information, the fact that not all nodes meet one another. 1-SW$^*$ goes beyond that and differentiates between neighboring nodes based on the quantitative expected inter-contact time. The fact that 1-SW$^*$ ourperforms the modified 1-SW thus indicates that routing actually benefits from the quantitative inter-contact time heterogeneity, and not just from node connectivity.

Table~\ref{res_simus} shows a descrepancy between the theoretical and the experimental delays.
This can be first explained by the presence of node pairs that do not have an exponential behavior. This is particularly true for 1-SW$^*$. In this case the computation of expected delays 
on mean inter-contact times misses possible inter-dependencies of node contacts. 

Simulation artifacts also come into play. The routing simulation is carried out on a limited time scale. 
The $\lambda$ values are computed over the entire data set in a prior pass, so a relay node may meet 
the destination for the last time before having met the source for the first 
time. This pre-computation being not realistic,
we could have used on-line predictive or learning methods.
However, as they are challenging to define, we let this study for future work 
and intend here to provide early validation results to motivate research 
in the domain.
The fact that 1-SW$^*$ delivers slightly less bundles 
than 1-SW is clearly due to this artefact. Indeed, because in 1-SW, the source
transfers more rapidly the bundle to a relay, we have a lower probability to 
contacts between that relay and the destination. 
Also, \textsc{Med} suffers from the same simulation artefact, 
it would have delivered $100$\% of bundles otherwise. 

Through these simulations, we validate the natural feeling that 
we should take into
account the heterogeneity of inter-contact time distributions in the
design of routing solutions for t-DTNs. Furthermore,
because 1-SW$^*$, which is only a two-hop protocol, achieves
better performance than \textsc{Med} by delivering bundles with lower delays 
and a lower impact on network resources (\textsc{Med} delivers bundles 
in 7.2 hops in average while 1-SW$^*$ uses only 2), we expect promising 
future work inspired from 1-SW$^*$ to be done. The opportunistic nature of 
1-SW$^*$ is the main reason of this superiority over \textsc{Med}, in which 
bundles follow a strict sequence of relays, in a network which is not 
a perfect exponential t-DTN.

\section{Conclusion}\label{sec:conclu}

We have first shown that, in a widely-used t-DTN data
set, distributions of inter-contact times are heterogeneous. 
As a consequence, one has to take it into account while modeling.
Second, we have validated the insight that considering heterogeneity
in routing improves performance. 
We presented a simple routing strategy, 1-SW$^*$,
adapted from the Spray and Wait scheme, which is capable of using 
only a subset of relays
to improve routing
performance, measured in term of average delay.

Clearly, our work, based as it is upon one data set, will benefit from
validation against others as mentioned in Sec.~\ref{sec:fitting}.
Work also needs to be done to examine why a memoryless model fits so 
many node pairs in an environment in which one would expect to find 
more temporal correlations. Finally, formal studies and validations
should be conducted with more elaborate schemes, in terms of number 
of copies distributed or in terms of the number of hops traversed.

\section*{Acknowledgments}

We gratefully acknowledge David Kotz for enabling our use of wireless
trace data from the CRAWDAD archive at Dartmouth College. We also thank 
Augustin Chaintreau for his valuable comments.

\bibliographystyle{splncs}

\begin{thebibliography}{10}

\bibitem{dtn_fall_sigcomm}
Fall, K.:
\newblock A delay-tolerant network architecture for challenged internets.
\newblock In: Proc. {SIGCOMM}. (2003)

\bibitem{UMassDieselNet}
{UM}ass{D}iesel{N}et:
\newblock {A} {B}us-based {D}isruption {T}olerant {N}etwork
  (\url{http://prisms.cs.umass.edu/diesel/})

\bibitem{chaintreau}
Chaintreau, A., Hui, P., Crowcroft, J., Diot, C., Gass, R., Scott, J.:
\newblock Impact of human mobility on the design of opportunistic forwarding
  algorithms.
\newblock In: Proc. {INFOCOM}. (2006)

\bibitem{spyro04single}
Spyropoulos, T., Psounis, K., Raghavendra, C.:
\newblock Single-copy routing in intermittently connected mobile networks.
\newblock In: Proc. {IEEE SECON}. (2004)

\bibitem{spyro04multi}
Spyropoulos, T., Psounis, K., Raghavendra, C.:
\newblock Multi-copy routing in intermittently connected mobile networks.
\newblock Technical report, USC (2004)

\bibitem{dartmount}
Henderson, T., Kotz, D., Abyzov, I.:
\newblock The changing usage of a mature campus-wide wireless network.
\newblock In: Proc. {M}obi{C}om. (2004)

\bibitem{song:predict}
Song, L., Kotz, D., Jain, R., He, X.:
\newblock Evaluating location predictors with extensive {Wi-Fi} mobility data.
\newblock In: Proc. {Infocom}. (2004)

\bibitem{leguay06}
Leguay, J., Friedman, T., Conan, V.:
\newblock Evaluating mobility pattern space routing for {DTN}s.
\newblock In: Proc. {INFOCOM}. (2006)

\bibitem{yoon06}
Yoon, J., Noble, B., Liu, M., Kim, M.:
\newblock Building realistic mobility models from coarse-grained traces.
\newblock In: Proc. {M}obi{S}ys. (2006)

\bibitem{jones_wdtn}
Jones, E.P.C., Li, L., Ward, P.A.S.:
\newblock Practical routing in delay-tolerant networks.
\newblock In: Proc. {WDTN}. (2005)

\bibitem{Eadie71}
Eadie, W.:
\newblock Statistical Methods in Experimental Physics.
\newblock Elsevier Science Ltd (1971)

\bibitem{realitymining}
Eagle, N., Pentland, A.:
\newblock Social serendipity: Mobilizing social software.
\newblock In: Proc. PerCom. (2005)

\bibitem{routingdtn}
Jain, S., Fall, K., Patra, R.:
\newblock Routing in a delay tolerant network.
\newblock In: Proc. {SIGCOMM}. (2004)

\bibitem{GrossglauserTse02}
Grossglauser, M., Tse, D.:
\newblock Mobility increases the capacity of ad-hoc wireless networks.
\newblock Transactions on Networking \textbf{10}(4) (2002)  477--486

\bibitem{Bremaud99}
Bremaud, P.:
\newblock Markov Chains, Gibbs Fields, Monte Carlo simulation, and queues.
\newblock Springer (1999)

\end{thebibliography}

\end{document}